\documentclass[letterpaper, 10 pt, conference]{ieeeconf}  

\IEEEoverridecommandlockouts                              

\usepackage{graphics} 
\usepackage{epsfig} 
\usepackage{mathptmx} 
\usepackage{times} 
\usepackage{amsmath} 
\usepackage{amssymb}  
\usepackage{algorithm}
\usepackage{mathrsfs}
\usepackage{cite}
\usepackage{graphicx} 
\usepackage{epstopdf}
\usepackage{subfig}
\usepackage{etoolbox}
\usepackage{algpseudocode}
\usepackage{algorithmicx}

\title{\LARGE \bf
Sybil-based Virtual Data Poisoning Attacks in Federated Learning*
}

\author{Changxun Zhu$^{1}$, Qilong Wu$^{1}$, Lingjuan Lyu$^{2}$ and Shibei Xue$^{1}$
\thanks{*This work was supported in part by the National Natural Science Foundation
 of China under Grant 62273226 and Grant 61873162. (Corresponding author:
 Shibei Xue)}
\thanks{$^{1}$Department of Automation, Shanghai Jiao Tong University, Shanghai 200240, P.R. China (e-mail: shbxue@sjtu.edu.cn).}%
\thanks{$^{2}$Sony AI}%
}

\begin{document}

\maketitle
\thispagestyle{empty}
\pagestyle{empty}

\begin{abstract}

Federated learning is vulnerable to poisoning attacks by malicious adversaries. Existing methods often involve high costs to achieve effective attacks. To address this challenge, we propose a sybil-based virtual data poisoning attack, where a malicious client generates sybil nodes to amplify the poisoning model's impact. To reduce neural network computational complexity, we develop a virtual data generation method based on gradient matching. We also design three schemes for target model acquisition, applicable to online local, online global, and offline scenarios. In simulation, our method outperforms other attack algorithms since our method can obtain a global target model under non-independent uniformly distributed data.
\end{abstract}


\section{INTRODUCTION}
The revolution in sensing technology has enabled high-quality data acquisition and processing across diverse real-world applications. This technological progress has catalyzed significant advancements in artificial intelligence (AI), achieving state-of-the-art performance in specialized domains including natural language processing\cite{devlin2019bert}, recommender systems\cite{guo2017deepfm,wang2017deep}, pose estimation\cite{guan2021high,guan2023hrpose}, intelligent transportation\cite{sheng2022graph,sheng2022cooperation}, energy-related prediction\cite{guan2020energy,Jia2019load,Liu2025prediction}.

However, with the growing emphasis on data privacy and the introduction of data protection regulations, traditional centralized machine learning approaches face significant obstacles\cite{lyu2020towards}. To address this, federated learning (FL) \cite{mcmahan2017communication} emerges as a privacy-preserving paradigm. FL establishes a shared model on a central server, distributes the model to clients for training on local data, and subsequently aggregates the locally trained models on the server. This framework avoids direct data transmission, thereby preserving client privacy \cite{kairouz2021advances}.

While federated learning preserves data locality on client devices, it also introduces new challenges. The inability to filter user data results in non-independent and identically distributed (Non-IID) data, leading to model drift and prolonged convergence times for optimal performance \cite{zhu2021federated}. Additionally, the lack of data filtering makes federated learning susceptible to attacks by malicious adversaries.

To address the aforementioned challenges, designing federated learning defense algorithms to enhance the stability of the federated learning process is a practical approach\cite{wu2022global}. Another perspective is to deepen the study of federated learning attack algorithms to understand potential security risks, thereby improving the security and privacy protection of federated learning. The latter can better understand the attack process from the attacker's perspective, which is more conducive to our formulation of proactive defense strategies.

The earliest poisoning attack was introduced against Support Vector Machines (SVM) by flipping the labels of training data \cite{biggio2012poisoning}. Although originally designed for centralized settings, this attack is found to be effective in federated learning scenarios \cite{tolpegin2020data}. Based on the structural characteristics of federated learning, the following poisoning attacks can be categorized into three types: data poisoning, model poisoning, and sybil-based poisoning attacks.

In data poisoning attacks, adversaries cannot directly manipulate users' models but can access and tamper with client training data to execute attacks. Ref. \cite{tolpegin2020data} first introduced label-flipping attacks to federated learning, where malicious actors flip sample labels, causing the trained model to deviate from the intended prediction boundary. However, the effectiveness of this approach is limited by the influence of non-malicious clients. To address this, Ref. \cite{shejwalkar2022back} proposed a dynamic label-flipping strategy that selects the target label with the smallest loss, improving on static label-flipping methods. Beyond label-flipping attacks, clean-label poisoning is another common data poisoning approach. This technique retains original labels but injects malicious patterns into model parameters through image pixel optimization\cite{biggio2012poisoning}. However, this approach is computationally expensive for deep neural networks. To overcome this limitation, heuristic methods have been proposed, as demonstrated in Refs. \cite{shafahi2018poison,geiping2020witches}, to achieve clean-label poisoning more efficiently.

However, the success rate of data poisoning attacks is directly proportional to the number of malicious clients controlled by the attacker, making such attacks costly in large-scale federated learning systems. To address this limitation, model poisoning attacks were introduced, enabling adversaries to manipulate the local training process. Ref. \cite{bagdasaryan2020backdoor} demonstrated an attack executed when the global model nears convergence, modifying the local training process by adding an anomaly detection term to the loss function. In contrast, Ref. \cite{bhagoji2019analyzing} proposed an attack targeting the early stages of global model training, before convergence is achieved. Additionally, Ref. \cite{zhou2021deep} leveraged a regularization term in the objective function to embed malicious neurons into the redundant spaces of neural networks. This approach minimizes the impact of benign clients during model aggregation, allowing the attacker to execute poisoning attacks effectively.

Sybil-based attacks are another common method of disrupting federated learning. In such attacks, a malicious attacker creates multiple fake clients to manipulate the model's learning process, potentially causing training errors or compromising private information. Ref. \cite{fung2020limitations} first introduced sybil-based attacks in federated learning, proposing a novel denial-of-service (DoS) attack. Inspired by this work, Refs. \cite{xiao2022sca, 9963704} employed sybil node collusion strategies to enhance attacker cooperation, enabling more effective poisoning attacks such as label flipping and backdoor attacks. However, the existing sybil-based poisoning attack methods often rely on sharing client data, which is costly and impractical in privacy-sensitive federated learning environments.

To address this challenge, we propose a sybil-based virtual data poisoning attack. Instead of sharing data with sybil nodes directly, we heuristically propose a novel approach for generating training data for sybil nodes based on gradient matching, which effectively reduces both the complexity of solving the optimization problem and the computational cost. Furthermore, we introduce three adaptive model acquisition strategies tailored to distinct deployment scenarios, enabling precise manipulation of model update directions. Experimental results demonstrate that our method significantly enhances the effectiveness of poisoning attacks compared to baseline methods.

The structure of this paper is as follows: Section II provides a brief overview of problem description and the sybil-based attack model in FL, followed by a detailed explanation of our proposed method in Section III. Section IV presents the experimental setup and results. Finally, Section V concludes the paper.

\section{PROBLEM DESCRIPTION OF SYBIL-BASED ATTACK}
In federated learning, data owners are referred to as clients, denoted by $\mathscr{C} = \{C_1, C_2, \dots, C_N\}$, where $N$ represents the total number of clients.
Each client $C_i$ has access to its local training dataset $D_i$. A central server oversees model initialization, aggregation, and distribution. The server begins by initializing a model $w_0$ and distributing it to all clients. Each client $C_i$ trains the model locally using its dataset $D_i$ and sends the updated model back to the server for aggregation.
\begin{figure}
\begin{center}
\includegraphics[width=0.4\textwidth]{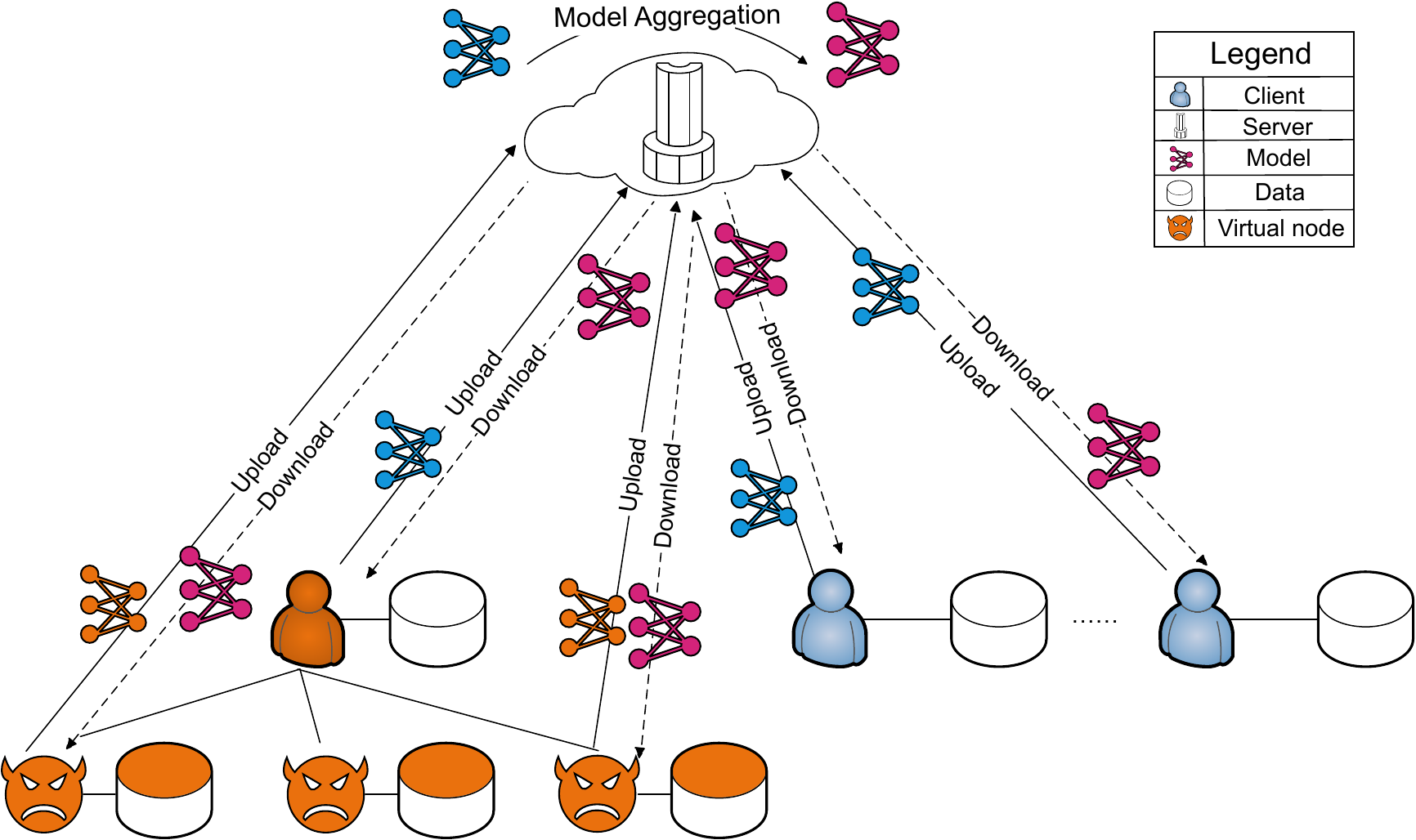}
\caption{Threat model for sybil-based attacks on FL.}
\label{fig1}
\end{center}
\end{figure}

The sybil-based attack model is illustrated in Fig. \ref{fig1}. In this model, attackers control a subset of malicious clients with real data to generate poison data. In a federated learning system with $N$ clients, we assume that the attacker can control $m\%$ of the clients as malicious and generate $v$ sybil nodes for each malicious client. The total number of sybil nodes, denoted as $\left|\mathscr{V}\right|$, is given by:
\begin{equation}
\left |\mathscr{V}\right |=N*m\%*v.
\label{eq0}
\end{equation}

To attack an image classification task, the objective is to train a model using poisoned data to misclassify images from a target category, while preserving the accuracy of classification for other categories of images.
Assume there are $N_1$ poisoning training samples from a target category and $N_2$ samples with normal labels. Taking the label-flipping attack as an example, we flip the labels of the target category from $y_i^{tar}$ to $y_i^{adv}$. Thus, our optimization problem can be formulated as follows:
\begin{equation}
    \min_{\varDelta \in \mathcal{C}}{\ \mathbf{J}=\displaystyle\sum_{i=1}^{{N}_{1}}}l\left ( {f}_{w\mathcal(\varDelta )}({x}_{i}),{y}_{i}^{adv}\right )+\displaystyle\sum_{i=1}^{{N}_{2}}l\left ( {f}_{w(\varDelta )}({x}_{i}),{y}_{i}\right ),
    \label{opo}
\end{equation}
\begin{equation}
s.t.\ w\left ( \varDelta \right )\in \arg \min_{w}\ \mathbf{\overline{J}}={\frac{1}{P}}\displaystyle\sum_{i=1}^{P}l\left ( {f}_{w}\left ( {x}_{i}+{\varDelta }_{i}\right ),{y}_{i}\right ),
\label{con1}
\end{equation}
\begin{equation}
C=\left \{ \varDelta \in {R}^{P\times n}: {\left \| \varDelta \right \|}_{\infty}\leq \epsilon \right \},
\label{con2}
\end{equation}
The first term in Eq. \eqref{opo} captures the error resulting from the misclassification of the target class, while the second term accounts for the error in correctly classifying images from other categories. ${f}_{w(\varDelta)}({x}_{i})$ represents the probability that the model $w(\varDelta)$ assigns a category prediction to $x_i$. Eq. \eqref{con1} defines the constraints for training with poisoned data on sybil nodes, where $P$ denotes the number of poisoned images and $\varDelta$ represents the perturbation introduced during the poisoning process. Eq. \eqref{con2} constrains the perturbations, with $C$ denoting the set of perturbations and $\epsilon$ indicating the perturbation threshold. $l$ representing the cross-entropy loss function, is written as
\begin{equation}
    l\left ( p,q\right )=\displaystyle\sum_{i=1}^{n}p\left ( {x}_{i}\right )\log{\frac{1}{q({x}_{i})}}=-\displaystyle\sum_{i=1}^{n}p\left ( {x}_{i}\right )\log{q\left ( {x}_{i}\right )}.
    \label{crossentropy}
\end{equation}
It is evident that our optimization problem forms a bilevel optimization structure. We refer to the objective function of the inner optimization, shown in Eq. \eqref{con1}, as the training loss, and the objective function of the outer optimization, shown in Eq. \eqref{opo}, as the adversarial loss.

\section{METHOD}

This section is divided into four subsections to detail the core components of the attack strategy. First, we introduce three approaches for obtaining the target model and a method for acquiring the baseline dataset. Next, to address the complexity of the task, we simplify the bilevel optimization problem using a gradient matching technique. After that, we describe the process of generating virtual poisoning data for sybil nodes to carry out poisoning attacks. Finally, we briefly introduce the overall algorithm workflow. 

\subsection{Acquisition of Target Model and Baseline Dataset}
In each round of the federated learning process, it is essential to determine the update direction of the poisoning model by obtaining the target model in advance. Additionally, we explain how to acquire a benchmark dataset for training malicious clients in this subsection.

To identify the target image with the label $y^{adv}$, we select images labeled as $y^{adv}$ from the controlled client's dataset as the baseline dataset $D_i^{base}$. Specifically, $D_i^{base}$ is defined as shown in Eq. \eqref{2-18}, where $D_i$ represents the local dataset of the controlled clients.
\begin{equation}
    {D}_{i}^{base}=\left \{ (x,y)\mid (x,y)\in {D}_{i},y={y}^{adv}\right \}.
    \label{2-18}
\end{equation}

To obtain the target model, we propose specific schemes for three scenarios: local online, global online, and offline.
In the online local target model acquisition scheme, the attacker performs a fake local training using the global model $w$ distributed by the server. The training data is defined in Eq. \eqref{2-21}. Using 
\begin{equation}
    {D}_{i}^{mdf}=\left  \{  ({x}_{i},{{y}_{i}}^{\prime})\mid ({x}_{i},{y}_{i})\in{D}_{i},{{y}_{i}}^{\prime}=\begin{cases}
{y}^{adv},& \mathrm{if}\, {y}_{i}={y}^{tar} \\
{y}_{i},& \mathrm{otherwise}
\end{cases}\right \},
\label{2-21}
\end{equation} 
the attacker trains a poisoning model $w_i^{mdf}$, which serves as the target model $w^{tar}$. Thus, ${w}^{tar} = {w}_{i}^{mdf}$.

Due to the non-IID nature of data across clients, obtaining the target model from a single client may hinder effective poisoning attacks. Some malicious clients may lack data for the target class, making label flipping infeasible. To address this, the online global target model scheme aggregates the $w_{i}^{mdf}$ models generated by all controlled malicious clients. The target model is defined as
\begin{equation}
    {w}^{tar}={w}^{mdf}=\displaystyle\sum_{i=1}^{M}\frac{1}{M}{w}_{i}^{mdf}.
    \label{2-22}
\end{equation}

Obtaining the target model during the poisoning process of federated learning demands excessive real-time communication resources. To address this, we propose an offline target model acquisition scheme which is implemented before federated learning begins.
The attacker distributes an untrained model to each controlled malicious client and uses their local label-flipped dataset $D_i^{mdf}$ for $R$ rounds of training and aggregation. The final aggregated model is used as the target model $w^{tar}$.

\subsection{Problem Simplification by Gradient Matching}
Given the complexity of our task, solving the optimization problem solely with neural networks is challenging. To address this, we simplify the calculation process using a gradient matching method.

In our bilevel optimization problem, the goal is to ensure that both the training and adversarial losses decrease concurrently through gradient descent. This allows the two objective functions to reach their low-value regions simultaneously. We thus have
\begin{flalign}
\frac{1}{{N}_{1}+{N}_{2}}\left ( \displaystyle\sum_{i=1}^{{N}_{1}}\nabla l\left ( {f}_{w}\left ( {x}_{i}\right ),{y}_{i}^{adv}\right )+\displaystyle\sum_{i=1}^{{N}_{1}}\nabla l\left ( {f}_{w}\left ( {x}_{i}\right ),{y}_{i}\right )\right )\notag
\\ \approx \frac{1}{P}\displaystyle\sum_{i=1}^{P}{\nabla}_{w} l\left ( {f}_{w}\left ( {x}_{i}+{\varDelta }_{i}\right ),{y}_{i}\right ).
\label{2-16}
\end{flalign}

However, finding poisoning images that satisfy Eq. \eqref{2-16} is challenging throughout the gradient descent process. In this case, we relax the requirement for gradient matching and instead aim to make the gradients of the model with respect to the objective functions of the inner and outer optimizations as similar as possible. Therefore, the bilevel optimization problem in Eqs. \eqref{opo}, \eqref{con1}, \eqref{con2} can be heuristically rewritten as
\begin{equation}
    \nabla \mathbf{J}=\textstyle\sum_{i=1}^{{N}_{1}}{\nabla }l\left ( {f}_{w}\left ( {x}_{i}\right ),{y}^{adv}_{i}\right )+\textstyle\sum_{i=1}^{{N}_{2}}{\nabla }l\left ( {f}_{w}\left ( {x}_{i}\right ),{y}_{i}\right ),
    \label{2-17_1}
\end{equation}
\begin{equation}
    \nabla \mathbf{\overline{J}} =\textstyle\sum_{i=1}^{P}{\nabla }_{w}l\left ( {f}_{w}\left ( {x}_{i}+{\varDelta }_{i}\right ),{y}_{i}\right ),
    \label{2-17_2}
\end{equation}
\begin{equation}
    \mathcal{B}\left ( \varDelta ;w\right )=1-\frac{\left \langle \nabla \mathbf{J} ,\nabla \mathbf{\overline{J}} \right \rangle}{\left \| \nabla \mathbf{J} \right \|\cdotp \left \|\nabla \mathbf{\overline{J}}  \right \|},
    \label{2-17_3}
\end{equation}
respectively.

We reformulate the original optimization problem using the form of negative cosine similarity, as shown in Eq. \eqref{2-17_3}, where $\left \| \cdot\right \|$ and $\left \langle \cdot\right \rangle$ represent the $L_{2}$ norm and dot product of vectors. By minimizing Eq. \eqref{2-17_3}, the value of the second term will gradually approach 1. According to the cosine similarity relationship, $\nabla\mathbf{J}$ and $\nabla\mathbf{\overline{J}}$ will become as similar as possible.
\subsection{Generation of Poisoning Data by Sybil Nodes}
To address the potential misalignment of gradient descent directions caused by mini-batch training in federated learning, we approximate the gradient using the negation of the difference between the target model $w^{tar}$ and the global model at the current round $w^{r}$. This approach provides a more representative descent direction as 
\begin{equation}
    \displaystyle\sum_{i=1}^{{N}_{1}}\nabla l({f}_{{w}^{r}}({x}_{i}),{y}_{i}^{adv})+\displaystyle\sum_{i=1}^{{N}_{2}}\nabla l({f}_{{w}^{r}}({x}_{i}),{y}_{i})\approx {w}^{r}-{w}^{tar}.
    \label{2-19}
\end{equation}
Consequently, Eq. \eqref{2-17_3} is reformulated as 
\begin{equation}
    \mathcal{B}(\varDelta ;{w}^{r})=1-\frac{\left \langle {w}^{r}-w^{tar},\textstyle\sum_{i=1}^{P}{\nabla }_{{w}^{r}}l({f}_{w^{r}}({x}_{i}+{\varDelta }_{i}),{y}_{i}) \right \rangle}{\left \| {w}^{r}-w^{tar}\right \|\cdotp \left \| \textstyle\sum_{i=1}^{P}{\nabla }_{{w}^{r}}l({f}_{w^{r}}({x}_{i}+{\varDelta }_{i}),{y}_{i})\right \|},
    \label{2-20}
\end{equation}
where $(x_{i},y_{i})$  are sampled from $D_{i}^{base}$.

As the local training process is immutable, the attacker can only introduce poisoned data through virtual nodes controlled by malicious clients. Using the target model $w^{tar}$ and the baseline dataset $D^{base}$, Eq. \eqref{2-20} is optimized via stochastic gradient descent, with the perturbation vector $\varDelta$ updated as
\begin{equation}
    \varDelta \left ( t+1\right )=\varDelta \left ( t\right )-{\nabla }_{\varDelta \left ( t\right )}\mathcal{B}.
    \label{2-23}
\end{equation}
The resulting poisoning data
\begin{equation}
    {D}_{poison}^{base}=\left \{ \left ( {x}^{\prime},y\right )\mid \left ( x,y\right )\in{D}_{i}^{base},{x}^{\prime}=x+{\varDelta }_{i}\right \}
    \label{x}
\end{equation}
is then injected into the sybil nodes.

\subsection{Complete Process}
In summary, our sybil-based poisoning attack algorithm is described in Algorithm $1$. In each training round, a subset $S_{r}$ is selected from the entire set of clients, including both real clients and virtual sybil nodes. Different training strategies are applied based on the client type within $S_{r}$. Benign clients perform standard local training, whereas malicious clients first obtain the target model and then generate poisoning data using their local dataset for training sybil nodes. The sybil nodes utilize the poisoning data generated by malicious clients for training and subsequently upload the poisoning models to the central server, where they contribute to the aggregation and update of the global model.
\renewcommand{\thealgorithm}{1}
    \begin{algorithm}
        \caption{Sybil-based poisoning attack algorithm}
        \begin{algorithmic}[1]
            \Require Communication rounds, $R$; The number of clients, $N$; Local training rounds, $E$; Learning rate, $\eta$;
            \Ensure Final model, ${w}^{R}$;
            \State Server execution:
            \State Initialize ${w}^{0}$
            \For{$r\gets 0$ \textbf{to} $R-1$}
                \State Select the client set ${S}_{r}$ from the $N+v*M$ clients
                \For{$i\gets1$ to $\left | {S}_{r}\right |$ parallel execution}
                    \State Send the global model ${w}^{r}$ to the client ${C}_{i}$
                    \If {$M \leq i \leq N$}
                    \State $w^{r}_{i} \gets$ Client local training $(i,w_{r})$
                    \ElsIf{$1 \leq i \leq M$}
                    \State ${w}^{tar} \gets$ Target model acquisition$(w^{r})$
                    \State $D_{poi}^{base}\gets$Poisoning data generation$(D_{i}^{base},w^{tar})$
                    \ElsIf{$1+N \leq i \leq N+v*M$}
                    \State Obtain $D_{i}$ from the malicious client
                    \State $w_{i}^{r} \gets$ Client local training $(i,w_{r})$
                    \EndIf
                \EndFor
                \State Server update: ${w}^{r+1}=\textstyle\sum_{i\in {S}_{r}}\frac{\left | D_{i}\right |}{{D}_{{S}_{r}}}{w}_{i}^{r}$
            \EndFor
            \State \Return $w^{R}$
            \State \textbf{Client local training} $(i,w_{r})$
            \State $w_{i}^{r}(0) \gets w^{r}$
            \For{$t \gets 0 $ \textbf{to} $\tau =\frac{\left | {D}_{i}\right |}{B}E-1$}
                \State ${l}_{i}\left ( {w}_{i}^{r}\left ( t\right )\right )\gets$ CrossEntropyLoss$\left ( {f}_{{w}_{i}^{r}(t)}\left ( x\right ),y\right )$
                \State ${w}_{i}^{r}\left ( t+1\right )\gets {w}_{i}^{r}\left ( t\right )-\eta \nabla {l}_{i}\left ( {w}_{i}^{r}\left ( t\right )\right )$
            \EndFor
            \State ${w}_{i}^{r} \gets {w}_{i}^{r}(\tau)$
            \State Send ${w}_{i}^{r}$ back to the server.
            \State \textbf{Poisoning data generation}
            \State $\varDelta \left ( 0\right ) \gets 0$
            \For{$t \gets 0$ \textbf{to} $T-1$}
                \State $\mathcal{B}\left ( \varDelta ;{w}^{r}\right )\gets $
                \Statex\qquad \qquad $1-\frac{\left \langle {w}^{tar}-{w}^{r},\textstyle\sum_{i=1}^{P}{\nabla }_{{w}^{r}}l({f}_{{w}^{r}}({x}_{i}+{\varDelta }_{i}),{y}_{i})\right \rangle}{\left \| {w}^{tar}-{w}^{r}\right \|\cdotp \left \|\textstyle\sum_{i=1}^{P}{\nabla }_{{w}^{r}}l({f}_{{w}^{r}}({x}_{i}+{\varDelta }_{i}),{y}_{i}) \right \|}$
                \State $\varDelta \left ( t+1\right ) \gets \varDelta \left ( t\right )-{\nabla }_{\varDelta \left ( t\right )}\mathcal{B}$
            \EndFor
            \State ${D}_{poison}^{base} \gets \left \{ \left ( {x}^{\prime},y\right )\mid \left ( x,y\right )\in {D}_{i}^{base},{x}^{\prime}=x+{\varDelta }_{i}\right \}$
            \State Send $D_{poison}^{base}$ to clients $C_{N+i},\cdotp \cdotp \cdotp ,C_{N+v}$
        \end{algorithmic}
    \end{algorithm}

\section{EXPERIMENT}
\subsection{Experimental Setup}
$\mathbf{Datasets.}$ To comprehensively evaluate the performance of our proposed algorithm on image classification tasks, we conduct experiments on three widely used datasets in federated learning: MNIST \cite{lecun1998gradient} (70K handwritten digit images, 10 classes), FMNIST \cite{xiao2017fashion} (70K fashion item images, 10 classes), and CIFAR-10 \cite{krizhevsky2009learning} (70K images, 10 classes). For MNIST and FMNIST, we use the online global target model acquisition scheme, while for CIFAR-10, we adopt the offline target model acquisition scheme.

$\mathbf{Networks.}$ We design training networks of varying complexity tailored to each dataset. For MNIST, a fully connected neural network (FC) with four layers is employed, the details of which are shown in Table \ref{tab2-1}. The final layer employs cross-entropy loss without activation for multi-classification. For FMNIST, a convolutional neural network (CNN) is employed. Details are provided in Table \ref{tab2-2}. A ReLU activation function \cite{glorot2011deep} follows each hidden layer. For CIFAR-10, the ResNet18 architecture is utilized.


\begin{table}[htbp]
\caption{Fully connected neural network structure table}
\begin{center}
\begin{tabular}{cccc}
\hline
Layers&Layer type&Input size&Output size\\
\hline
1&FC&[784]&[32]\\
\hline
2&FC&[32]&[16]\\
\hline
3&FC&[16]&[8]\\
\hline
4&FC&[8]&[10]\\
\hline
\end{tabular}
\label{tab2-1}
\end{center}
\end{table}
\begin{table}[htbp]
\caption{Convolutional neural network structure table}
\begin{center}
\begin{tabular}{cccc}
\hline
Layers&Layer type&Input size&Output size\\
\hline
1&CONV&[1,28,28]&[6,24,24]\\
\hline
2&MaxPool&[6,24,24]&[6,12,12]\\
\hline
3&CONV&[6.12,12]&16,8,8]\\
\hline
4&MaxPool&[16,8,8]&[16,4,4]\\
\hline
5&FC&[16,4,4]&[120]\\
\hline
6&FC&[120]&[84]\\
\hline
7&FC&[84]&[10]\\
\hline
\end{tabular}
\label{tab2-2}
\end{center}
\end{table}

$\mathbf{Environment\ setting.}$ We implement the federated learning process using PyTorch (version 1.13.0) in a distributed training setup. For local training on clients, we employ the SGD optimizer with a learning rate of $\eta = 0.01$ and momentum of 0.9. Each client train for $E = 5$ cycles per round with a batch size of $B = 64$. The total number of clients $N = 50$, all of which participate in every communication round. The number of communication rounds is set to $R = 300$ for MNIST and FMNIST, and $R = 200$ for CIFAR-10, at which point the federated averaging algorithm achieve stable accuracy. Based on experience, attacks are executed during the final 50 rounds for all three datasets. Simulations were conducted using an Intel Xeon Platinum 8255C CPU, 40 GB RAM, and four NVIDIA RTX 3080 GPUs on a single machine configuration.

$\mathbf{Parameters.}$ By default, the number of malicious clients is $M = 20$ (40\% of the total). Each client generates $v = 5$ sybil nodes, and the Dirichlet distribution hyperparameter is set to $\alpha = 0.5$. The disturbance vector $\varDelta$ is unconstrained in size, and poisoning data generation halts after a fixed number of iterations, set to $T = 300$, with a learning rate of 1. For each malicious client, we compute the disturbance vector for only 32 images from its baseline data. By default, the goal is to misclassify data labeled as `1' into `7', making the target category 1, the adversarial category 7, and the baseline category also 7.

$\mathbf{Evaluation\ Metrics.}$ This study employs Main Task Accuracy (MTA) and Target Task Accuracy (TTA) as evaluation metrics. In our study, TTA measures the accuracy of the global model in classifying target images into the desired category. MTA evaluates the accuracy of the global model in classifying non-target images into their original categories. Both of these metrics are evaluated under poisoning attacks. Additionally, we compare MTA with the global model accuracy (GMA) which is computed in the absence of malicious clients.

\subsection{Effect of the Number of Sybil Nodes}
\begin{figure}[htbp]
  \centering
  \subfloat[MNIST dataset]{\includegraphics[width=0.15\textwidth]{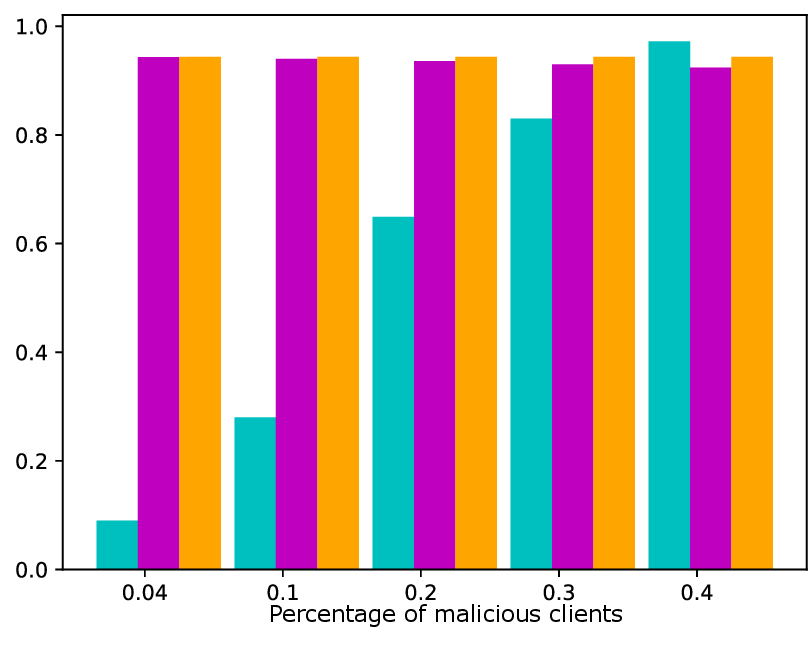}}
  \hfill
  \subfloat[FMNIST dataset]{\includegraphics[width=0.15\textwidth]{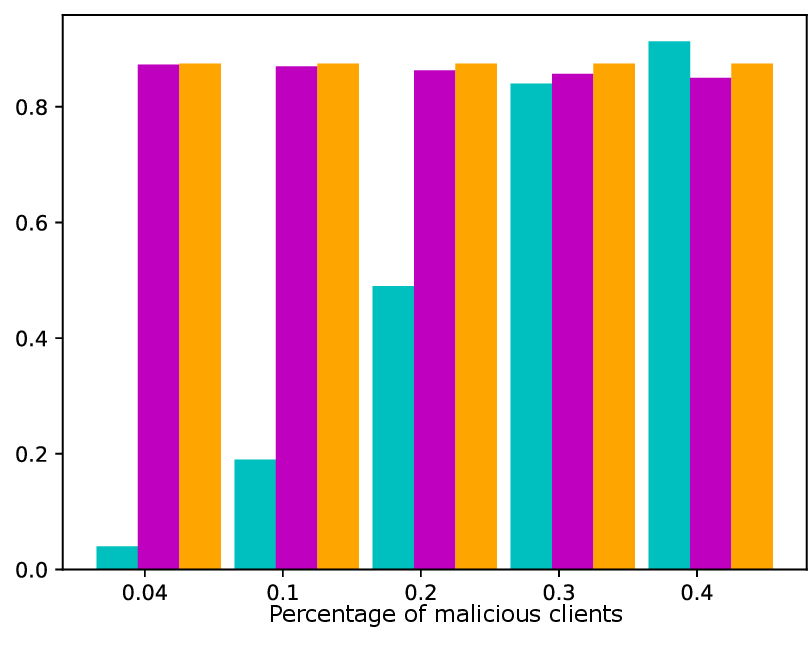}}
  \hfill
  \subfloat[CIFAR-10 dataset]{\includegraphics[width=0.15\textwidth]{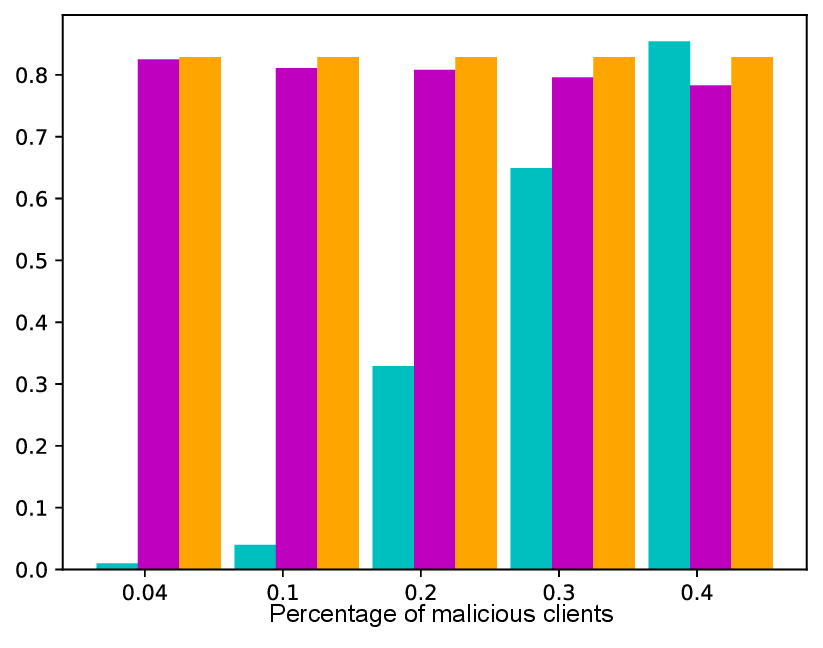}}
  \hfill
  \vspace{2pt}
  \subfloat{\includegraphics[width=0.4\textwidth]{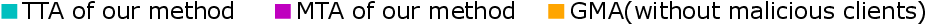}}
  \caption{The test accuracy changes with the proportion of malicious clients while fix the number of virtual nodes generated by each client $v$ = 5.}
  \label{fig5}
\end{figure}

\begin{figure}[htbp]
  \centering
  \subfloat[MNIST dataset]{\includegraphics[width=0.15\textwidth]{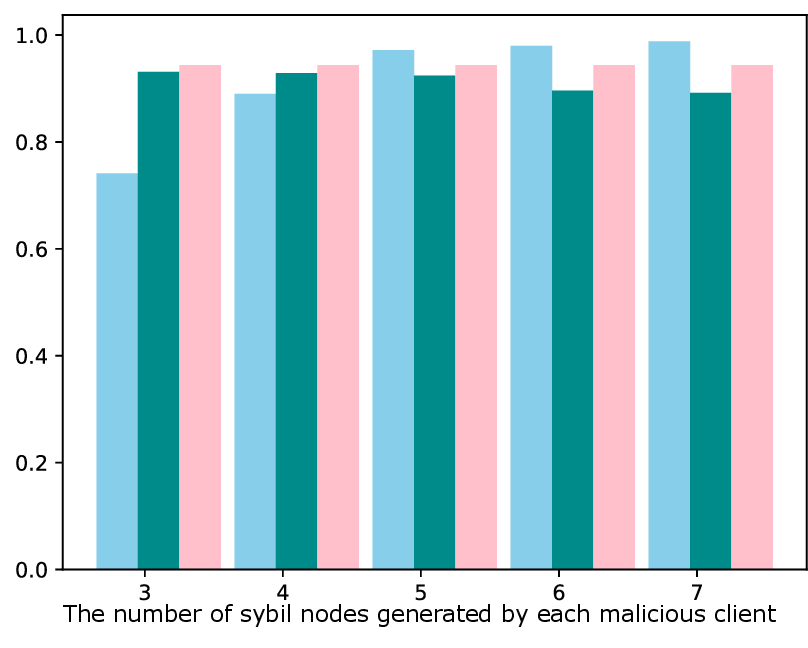}}
  \hfill
  \subfloat[FMNIST dataset]{\includegraphics[width=0.15\textwidth]{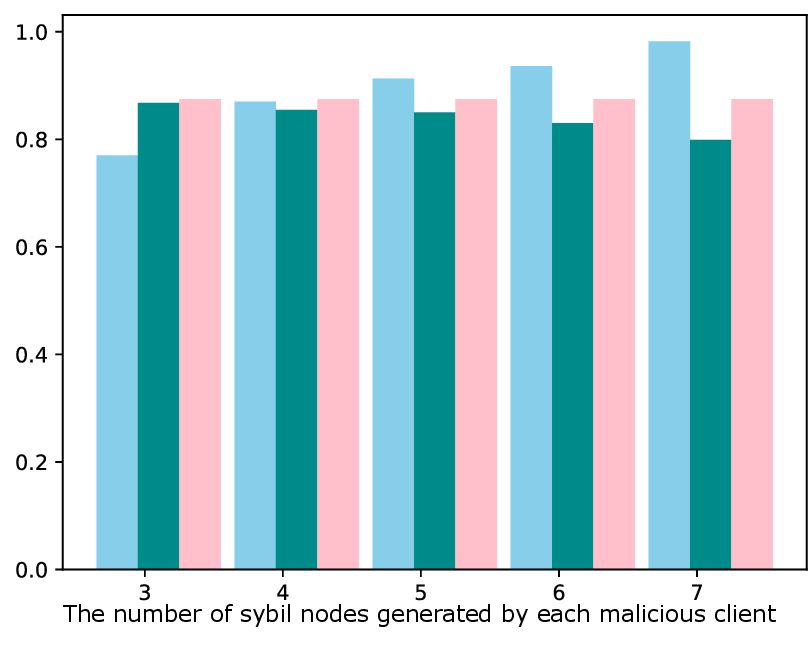}}
  \hfill
  \subfloat[CIFAR-10 dataset]{\includegraphics[width=0.15\textwidth]{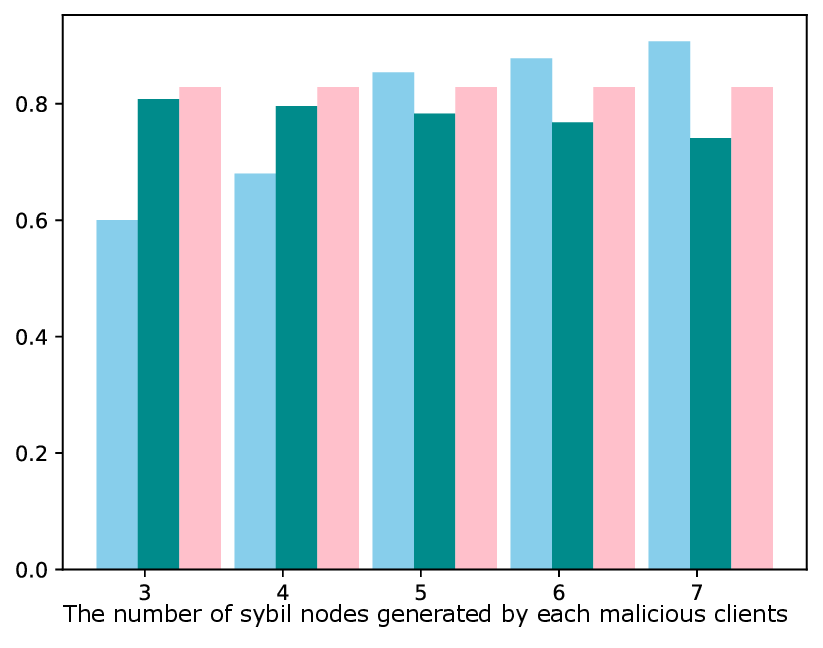}}
  \hfill
  \vspace{2pt}
  \subfloat{\includegraphics[width=0.4\textwidth]{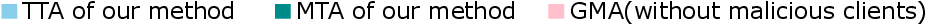}}
  \caption{The test accuracy changes with the number of virtual nodes generated by each malicious clients while fix the proportion of malicious clients to 0.4.}
  \label{fig6}
\end{figure}

This experiment investigates the impact of the proportion of malicious clients and the number of sybil nodes generated by each on the poisoning attack. First, we fix the proportion of malicious clients at $m\%=40\%$ and vary the number of sybil nodes generated by each malicious client $v = \{5, 6, 7, 8, 9\}$. After that, we fix $v = 5$ and vary the proportion of malicious clients $m\% = \{4\%, 10\%, 20\%, 30\%, 40\%\}$. In both cases, attacks are launched in the last 50 rounds across all three datasets. The results are shown in Fig.\ref{fig5} and Fig.\ref{fig6}.

The results in the figures show that, across all three datasets, increasing either
$m\%$ and $v$ improves the attacker’s target task precision on the test dataset. Notably, increasing the proportion of malicious clients has a more immediate impact on attack performance, as it provides more client data for computing the target model. However, both increasing the proportion of malicious clients and increasing the number of sybil nodes generated by each malicious client reduce the accuracy of the main task, although the overall decrease is modest. Thus, balancing target task accuracy and main task accuracy, we set $m\%=40\%$ and $v=5$ for subsequent experiments.
\subsection{Comparative Experiment}
We implement the method from \cite{shafahi2018poison} within our sybil-based data poisoning framework. In this methode, malicious clients generate poisoning data using the global model and sends it to the sybil nodes for poisoning. This method is referred as the Feature Collision Method (FCM). The approach in \cite{geiping2020witches} is similar to the online local target model scheme we proposed, with the key difference being that it only considers the target class of images in the counter loss, excluding benign samples from other classes. We refer to this method as the Local Method (LM).

\begin{figure}[htbp]
  \centering
  \subfloat[MNIST dataset]{\includegraphics[width=0.16\textwidth]{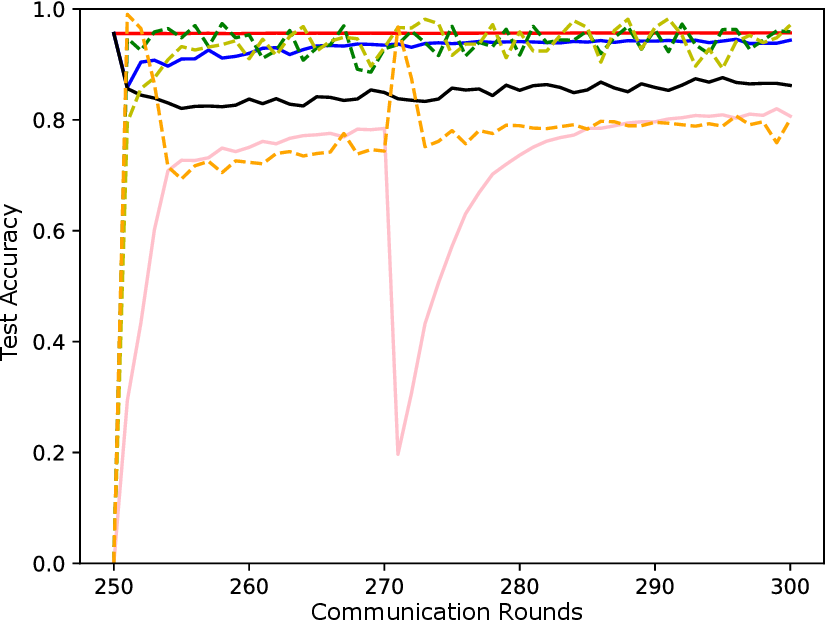}}
  \hfill
  \subfloat[FMNIST dataset]{\includegraphics[width=0.16\textwidth]{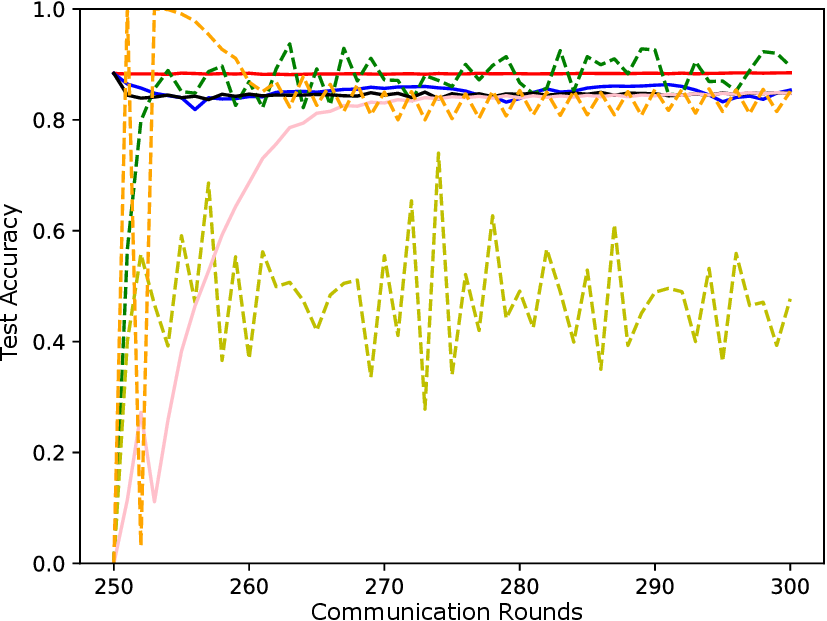}}
  \hfill
  \subfloat[CIFAR-10 dataset]{\includegraphics[width=0.16\textwidth]{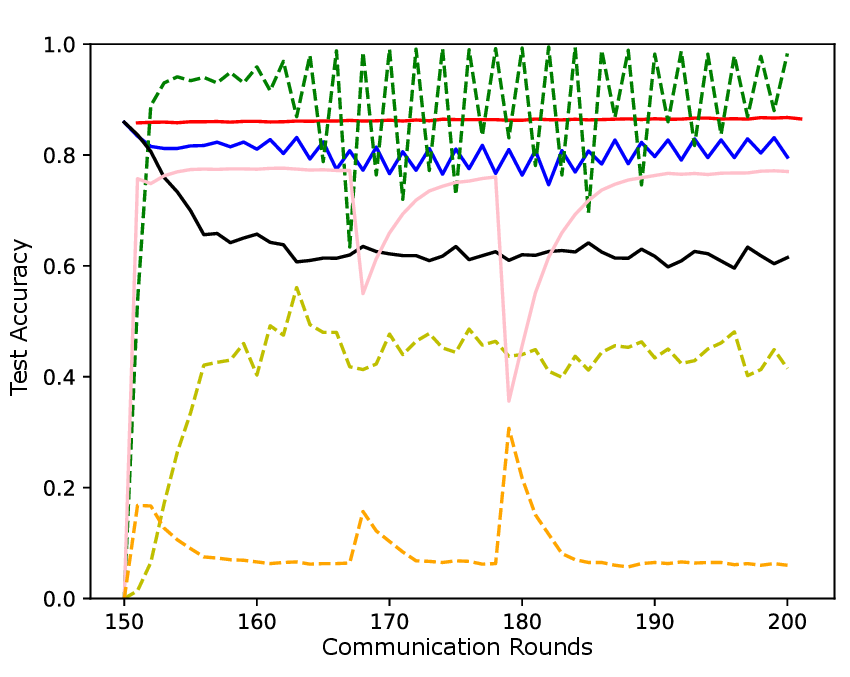}}
  \hfill
  \vspace{2pt}
  \subfloat{\includegraphics[width=0.4\textwidth]{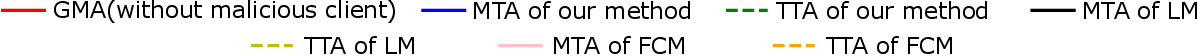}}
  \caption{Attack performance when three datasets are IID.}
  \label{fig7}
\end{figure}

\begin{figure}[htbp]
  \centering
  \subfloat[MNIST dataset]{\includegraphics[width=0.16\textwidth]{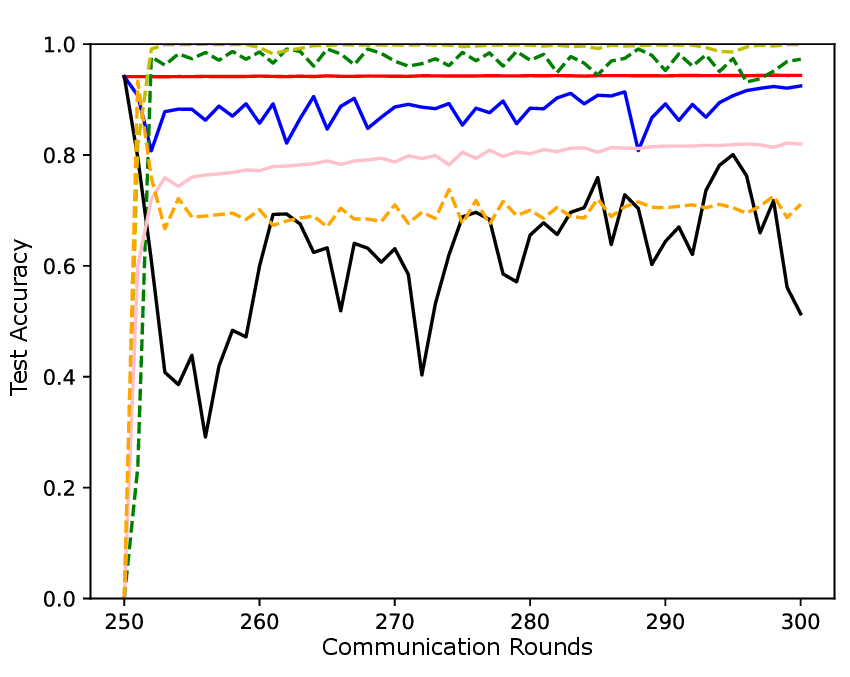}}
  \hfill
  \subfloat[FMNIST dataset]{\includegraphics[width=0.16\textwidth]{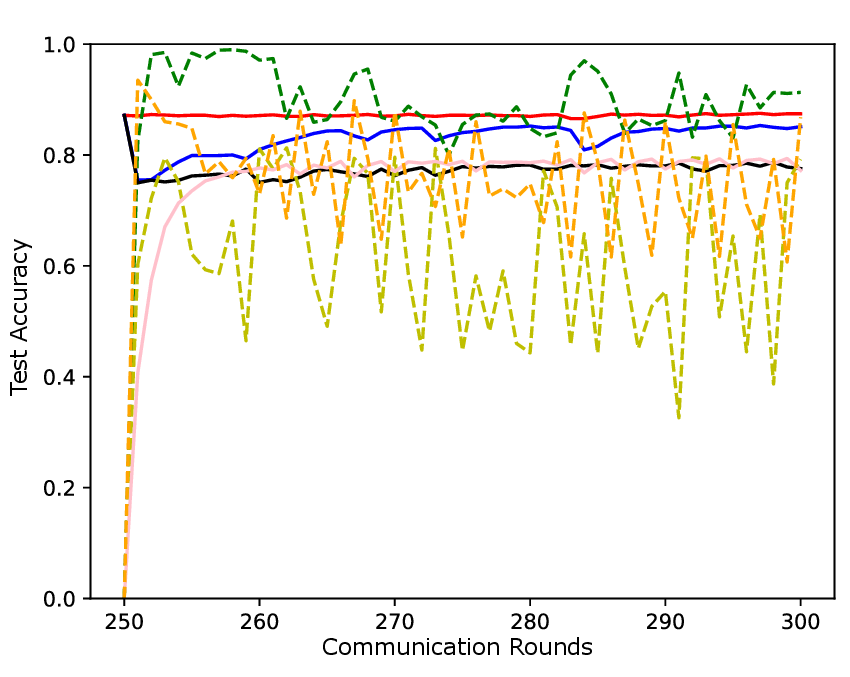}}
  \hfill
  \subfloat[CIFAR-10 dataset]{\includegraphics[width=0.16\textwidth]{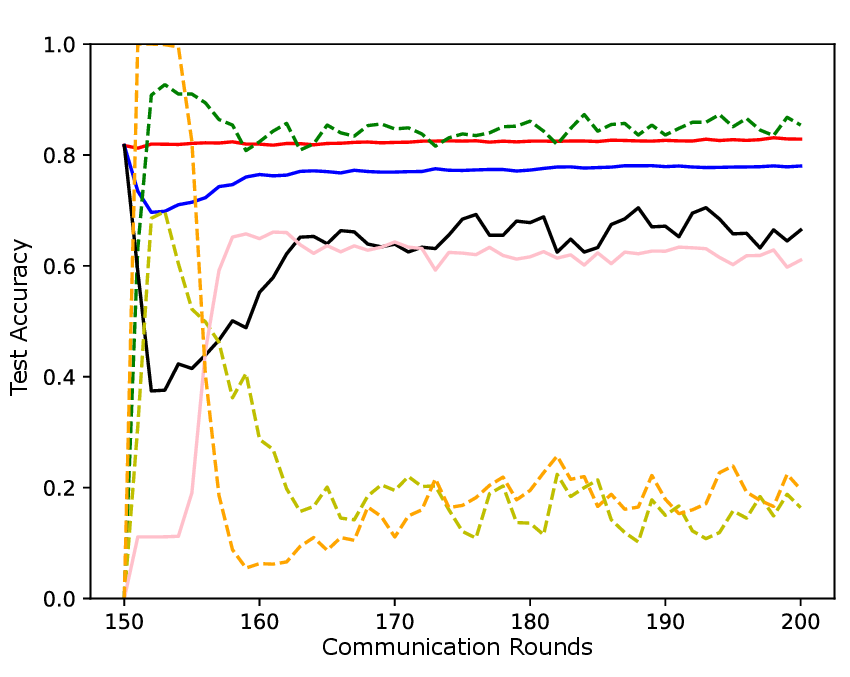}}
  \hfill
  \vspace{2pt}
  \subfloat{\includegraphics[width=0.4\textwidth]{fig/atkcaption.eps}}
  \caption{Attack performance when the Dirichlet distribution has parameter $\alpha$ = 0.5.}
  \label{fig8}
\end{figure}

\begin{figure}[htbp]
  \centering
  \subfloat[MNIST dataset]{\includegraphics[width=0.16\textwidth]{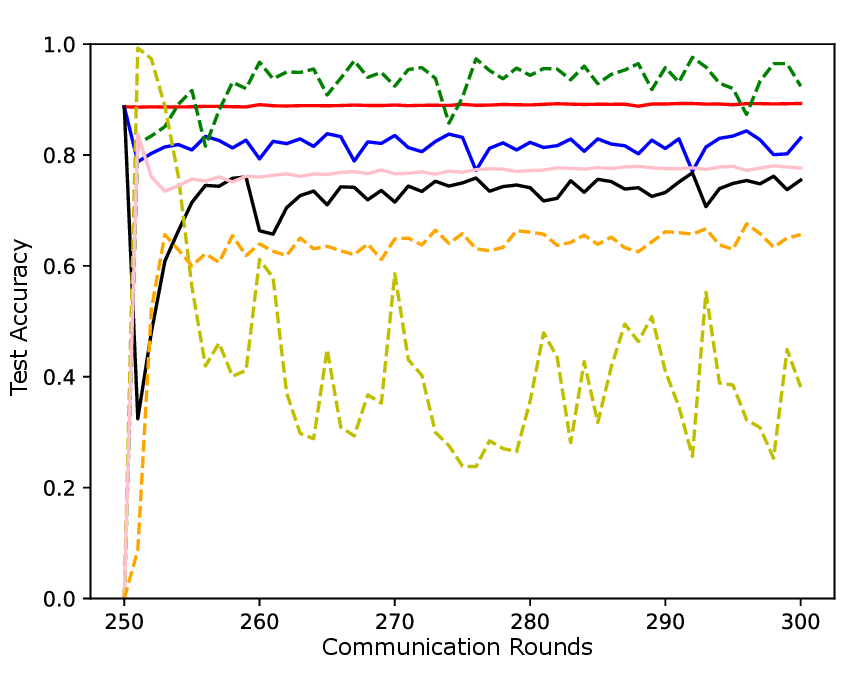}}
  \hfill
  \subfloat[FMNIST dataset]{\includegraphics[width=0.16\textwidth]{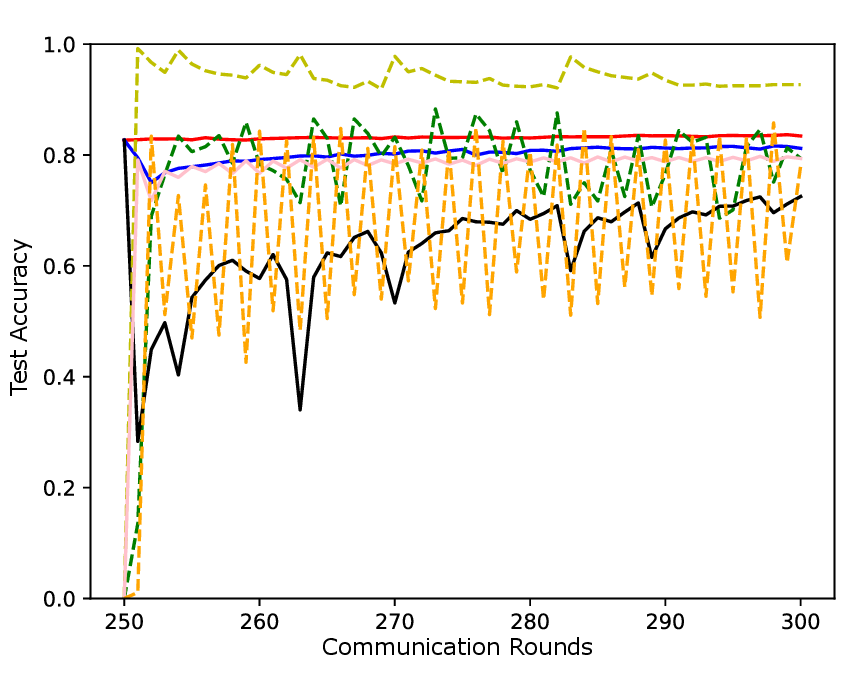}}
  \hfill
  \subfloat[CIFAR-10 dataset]{\includegraphics[width=0.16\textwidth]{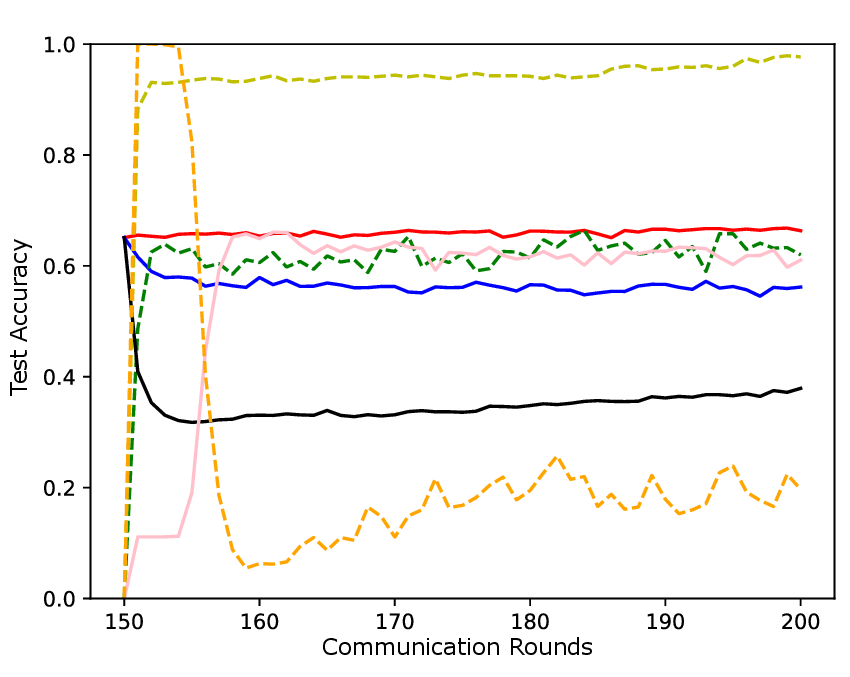}}
  \hfill
  \vspace{2pt}
  \subfloat{\includegraphics[width=0.4\textwidth]{fig/atkcaption.eps}}
  \caption{Attack performance when the Dirichlet distribution has parameter $\alpha$ = 0.1.}
  \label{fig9}
\end{figure}

Non-independent, identically distributed (Non-IID) data is a key characteristic of federated learning, and varying heterogeneous data distributions can influence the effectiveness of poisoning attacks. Using the two methods described above and our proposed method, we conducted comparative experiments with three data distributions: IID, Dirichlet distribution with $\alpha = 0.5$, and Dirichlet distribution with $\alpha = 0.1$. The experimental results are shown in Fig.\ref{fig7}, Fig.\ref{fig8}, and Fig.\ref{fig9}.

Overall, the method proposed in this paper outperforms others across all three datasets in both main task accuracy (MTA) and target task accuracy (TTA). Specifically, the poisoning data we generate is highly dependent on the target model, and its quality is influenced by the dataset held by the controlled malicious clients. Even when using the Dirichlet distribution with $\alpha = 0.1$, which simulates highly non-independent data, the poisoning attack remains effective. On these three datasets, the TTA of the proposed method is 92.42\%, 80.43\%, and 63.74\%, respectively. Although the TTA is not the highest—being 12.27\% and 33.96\% lower than the LM method for FMNIST and CIFAR-10—the MTA of LM has significantly decreased, rendering the poisoning attack less effective. The experimental results show that the data poisoning attack method proposed in this paper plays a significant role in improving the performance of the poisoning attack by considering non-target samples in the objective function and obtaining the global target model on the non-IID data.

\section{conclusion}
In this paper, we have proposed a sybil-based virtual data poisoning attack that leverages sybil nodes to amplify the impact of the poisoning attack while minimizing the high cost associated with malicious clients directly providing data. Our method first computes the target model, then generates virtual data on the malicious client, which is distributed to the corresponding sybil node for federated learning participation, thereby poisoning the global model. In experiments, we explore the appropriate proportion of malicious clients and the number of sybil nodes using non-IID datasets. We have also compared our approach with existing algorithms under various data distributions. Compared to the second-best local method, the main task accuracy improved by 7.6\%, 9.03\%, and 17.3\%. 


\bibliographystyle{IEEEtran}
\bibliography{IEEEabrv,ifacconf.bib}%

\begin{thebibliography}{10}
\providecommand{\url}[1]{#1}
\csname url@samestyle\endcsname
\providecommand{\newblock}{\relax}
\providecommand{\bibinfo}[2]{#2}
\providecommand{\BIBentrySTDinterwordspacing}{\spaceskip=0pt\relax}
\providecommand{\BIBentryALTinterwordstretchfactor}{4}
\providecommand{\BIBentryALTinterwordspacing}{\spaceskip=\fontdimen2\font plus
\BIBentryALTinterwordstretchfactor\fontdimen3\font minus
  \fontdimen4\font\relax}
\providecommand{\BIBforeignlanguage}[2]{{%
\expandafter\ifx\csname l@#1\endcsname\relax
\typeout{** WARNING: IEEEtran.bst: No hyphenation pattern has been}%
\typeout{** loaded for the language `#1'. Using the pattern for}%
\typeout{** the default language instead.}%
\else
\language=\csname l@#1\endcsname
\fi
#2}}
\providecommand{\BIBdecl}{\relax}
\BIBdecl

\bibitem{devlin2019bert}
J.~Devlin, M.-W. Chang, K.~Lee, and K.~Toutanova, ``Bert: Pre-training of deep
  bidirectional transformers for language understanding,'' in \emph{Proceedings
  of the 2019 conference of the North American chapter of the association for
  computational linguistics: human language technologies, volume 1 (long and
  short papers)}, 2019, pp. 4171--4186.

\bibitem{guo2017deepfm}
H.~Guo, R.~Tang, Y.~Ye, Z.~Li, and X.~He, ``Deepfm: a factorization-machine
  based neural network for ctr prediction,'' \emph{arXiv preprint
  arXiv:1703.04247}, 2017.

\bibitem{wang2017deep}
R.~Wang, B.~Fu, G.~Fu, and M.~Wang, ``Deep \& cross network for ad click
  predictions,'' in \emph{Proceedings of the ADKDD'17}.\hskip 1em plus 0.5em
  minus 0.4em\relax New York, NY, USA: Association for Computing Machinery,
  2017, pp. 1--7.

\bibitem{guan2021high}
Q.~Guan, W.~Li, S.~Xue, and D.~Li, ``High-resolution representation object pose
  estimation from monocular images,'' in \emph{2021 China Automation Congress
  (CAC)}.\hskip 1em plus 0.5em minus 0.4em\relax IEEE, 2021, pp. 980--984.

\bibitem{guan2023hrpose}
Q.~Guan, Z.~Sheng, and S.~Xue, ``Hrpose: Real-time high-resolution 6d pose
  estimation network using knowledge distillation,'' \emph{Chinese Journal of
  Electronics}, vol.~32, no.~1, pp. 189--198, 2023.

\bibitem{sheng2022graph}
Z.~Sheng, Y.~Xu, S.~Xue, and D.~Li, ``Graph-based spatial-temporal
  convolutional network for vehicle trajectory prediction in autonomous
  driving,'' \emph{IEEE Transactions on Intelligent Transportation Systems},
  vol.~23, no.~10, pp. 17\,654--17\,665, 2022.

\bibitem{sheng2022cooperation}
Z.~Sheng, L.~Liu, S.~Xue, D.~Zhao, M.~Jiang, and D.~Li, ``A cooperation-aware
  lane change method for automated vehicles,'' \emph{IEEE Transactions on
  Intelligent Transportation Systems}, vol.~24, no.~3, pp. 3236--3251, 2022.

\bibitem{guan2020energy}
Y.~Guan, D.~Li, S.~Xue, and Y.~Xi, ``Feature-fusion-kernel-based gaussian
  process model for probabilistic long-term load forecasting,''
  \emph{Neurocomputing}, vol. 426, pp. 174--184, 2020.

\bibitem{Jia2019load}
S.~Jia, Z.~Gan, Y.~Xi, D.~Li, S.~Xue, and L.~Wang, ``A deep reinforcement
  learning bidding algorithm on electricity market,'' \emph{Journal of Thermal
  Science}, vol.~29, no.~5, pp. 1125--1134, 2020.

\bibitem{Liu2025prediction}
L.~Liu, J.~Zhang, and S.~Xue, ``Photovoltaic power forecasting: Using wavelet
  threshold denoising combined with vmd,'' \emph{Renewable Energy}, vol. 249,
  p. 123152, 2025.

\bibitem{lyu2020towards}
L.~Lyu, Y.~W. Law, K.~S. Ng, S.~Xue, J.~Zhao, M.~Yang, and L.~Liu, ``Towards
  distributed privacy-preserving prediction,'' in \emph{2020 IEEE International
  Conference on Systems, Man, and Cybernetics (SMC)}.\hskip 1em plus 0.5em
  minus 0.4em\relax IEEE, 2020, pp. 4179--4184.

\bibitem{mcmahan2017communication}
B.~McMahan, E.~Moore, D.~Ramage, S.~Hampson, and B.~A. y~Arcas,
  ``Communication-efficient learning of deep networks from decentralized
  data,'' in \emph{Proceedings of the 20th International Conference on
  Artificial Intelligence and Statistics}, 2017, pp. 1273--1282.

\bibitem{kairouz2021advances}
P.~Kairouz, H.~B. McMahan, B.~Avent, A.~Bellet, M.~Bennis, A.~N. Bhagoji,
  K.~Bonawitz, Z.~Charles, G.~Cormode, R.~Cummings \emph{et~al.}, ``Advances
  and open problems in federated learning,'' \emph{Foundations and
  trends{\textregistered} in machine learning}, vol.~14, no. 1--2, pp. 1--210,
  2021.

\bibitem{zhu2021federated}
H.~Zhu, J.~Xu, S.~Liu, and Y.~Jin, ``Federated learning on non-iid data: A
  survey,'' \emph{Neurocomputing}, vol. 465, pp. 371--390, 2021.

\bibitem{wu2022global}
Q.~Wu, L.~Liu, and S.~Xue, ``Global update guided federated learning,'' in
  \emph{2022 41st Chinese Control Conference (CCC)}.\hskip 1em plus 0.5em minus
  0.4em\relax IEEE, 2022, pp. 2434--2439.

\bibitem{biggio2012poisoning}
B.~Biggio, B.~Nelson, and P.~Laskov, ``Poisoning attacks against support vector
  machines,'' \emph{arXiv:1206.6389}, 2012.

\bibitem{tolpegin2020data}
V.~Tolpegin, S.~Truex, M.~E. Gursoy, and L.~Liu, ``Data poisoning attacks
  against federated learning systems,'' in \emph{25th European symposium on
  research in computer security}, 2020, pp. 480--501.

\bibitem{shejwalkar2022back}
V.~Shejwalkar, A.~Houmansadr, P.~Kairouz, and D.~Ramage, ``Back to the drawing
  board: A critical evaluation of poisoning attacks on production federated
  learning,'' in \emph{IEEE Journal on Emerging and Selected Topics in Circuits
  and Systems}, 2022, pp. 1354--1371.

\bibitem{shafahi2018poison}
A.~Shafahi, W.~R. Huang, M.~Najibi, O.~Suciu, C.~Studer, T.~Dumitras, and
  T.~Goldstein, ``Poison frogs! targeted clean-label poisoning attacks on
  neural networks,'' \emph{Advances in neural information processing systems},
  vol.~31, pp. 6106--6116, 2018.

\bibitem{geiping2020witches}
J.~Geiping, L.~Fowl, W.~R. Huang, W.~Czaja, G.~Taylor, M.~Moeller, and
  T.~Goldstein, ``Witches' brew: Industrial scale data poisoning via gradient
  matching,'' \emph{arXiv:2009.02276}, 2020.

\bibitem{bagdasaryan2020backdoor}
E.~Bagdasaryan, A.~Veit, Y.~Hua, D.~Estrin, and V.~Shmatikov, ``How to backdoor
  federated learning,'' in \emph{International conference on artificial
  intelligence and statistics}, 2020, pp. 2938--2948.

\bibitem{bhagoji2019analyzing}
A.~N. Bhagoji, S.~Chakraborty, P.~Mittal, and S.~Calo, ``Analyzing federated
  learning through an adversarial lens,'' in \emph{International conference on
  machine learning}, 2019, pp. 634--643.

\bibitem{zhou2021deep}
X.~Zhou, M.~Xu, Y.~Wu, and N.~Zheng, ``Deep model poisoning attack on federated
  learning,'' \emph{Future Internet}, vol.~13, no.~3, pp. 73--88, 2021.

\bibitem{fung2020limitations}
C.~Fung, C.~J. Yoon, and I.~Beschastnikh, ``The limitations of federated
  learning in sybil settings,'' in \emph{23rd International Symposium on
  Research in Attacks, Intrusions and Defenses}, 2020, pp. 301--316.

\bibitem{xiao2022sca}
X.~Xiao, Z.~Tang, C.~Li, B.~Xiao, and K.~Li, ``Sca: Sybil-based collusion
  attacks of iiot data poisoning in federated learning,'' \emph{IEEE
  Transactions on Industrial Informatics}, vol.~19, no.~3, pp. 2608--2618,
  2022.

\bibitem{9963704}
X.~Xiao, Z.~Tang, C.~Li, B.~Jiang, and K.~Li, ``Sbpa: Sybil-based backdoor
  poisoning attacks for distributed big data in aiot-based federated learning
  system,'' \emph{IEEE Transactions on Big Data}, vol.~10, no.~6, pp. 827--838,
  2024.

\bibitem{lecun1998gradient}
Y.~LeCun, L.~Bottou, Y.~Bengio, and P.~Haffner, ``Gradient-based learning
  applied to document recognition,'' \emph{Proceedings of the IEEE}, vol.~86,
  no.~11, pp. 2278--2324, 1998.

\bibitem{xiao2017fashion}
H.~Xiao, K.~Rasul, and R.~Vollgraf, ``Fashion-mnist: a novel image dataset for
  benchmarking machine learning algorithms,'' \emph{arXiv:1708.07747}, 2017.

\bibitem{krizhevsky2009learning}
A.~Krizhevsky, G.~Hinton \emph{et~al.}, ``Learning multiple layers of features
  from tiny images,'' \emph{Technical report, University of Toronto}, pp.
  1--58, 2009.

\bibitem{glorot2011deep}
X.~Glorot, A.~Bordes, and Y.~Bengio, ``Deep sparse rectifier neural networks,''
  in \emph{Proceedings of the fourteenth international conference on artificial
  intelligence and statistics}, 2011, pp. 315--323.

\end{thebibliography}

\addtolength{\textheight}{-12cm}   




\end{document}